\documentclass{article}

\usepackage{arxiv}

\usepackage{hyperref}       
\usepackage{url}            
\usepackage{booktabs}       
\usepackage{amsfonts}       
\usepackage{nicefrac}       
\usepackage{microtype}      
\usepackage{lipsum}

\usepackage{graphicx}
\usepackage{float}
\usepackage{amsmath}
\usepackage{subfigure}
\usepackage{lineno}
\usepackage[noend]{algpseudocode}
\usepackage{algorithm}

\usepackage{amssymb}
\usepackage{xcolor}

\usepackage{subcaption}

\title{A physics-inspired nonlinear momentum method for gradient descent with applications to inverse photonic design}

\author{
  Jianing Zhang$^{1, \ast}$ \\
  Northeastern University, China \\
  \And
  Rumei Liu$^{2, \ast}$ \\
  Liaoning University, China \\
  \And
  \vspace{-2em} \\ %
  \small $^\ast$Corresponding authors: zhangjn@neu.edu.cn (J. Zhang), liurumei@lnu.edu.cn (R. Liu)
}

\begin{document}
\maketitle


\begin{abstract}  
In this work, a nonlinear momentum method is introduced to enhance the convergence performance of momentum-based gradient optimization algorithms. Classical momentum methods, such as the Heavy Ball method, can be viewed as a dynamical system with quadratic kinetic energy and linear damping. By extending this analogy to non-Newtonian dynamical systems, we construct a Hamiltonian framework for optimization problems. In this framework, nonlinear kinetic energy and nonlinear damping effects naturally emerge. It provides a more flexible and physically interpretable mechanism for optimization algorithms. Specifically, we employ an anharmonic kinetic energy function to capture the inertial effects of accumulated gradient information during the optimization process, while the nonlinear damping mechanism effectively regulates the contribution of momentum during convergence. Numerical experiments show that the proposed method achieves faster convergence compared to classical momentum algorithms, making it particularly suitable for inverse design tasks. Moreover, the Hamiltonian based algorithmic framework may offer physical insights for the development of efficient physics-inspired optimization algorithms.
\end{abstract}

\section{Introduction}
Physics has long served as a fundamental source of inspiration for numerical optimization. Classical examples include Newton-type and quasi-Newton methods \cite{Nocedal06}, which were originally derived from physical models of force balance and dynamical relaxation. Momentum-based optimization methods, such as Polyak’s Heavy Ball (HB) method \cite{POLYAK64}, can be interpreted as damped dynamical systems. Modern algorithms—like simulated annealing \cite{kirk83}, Langevin dynamics \cite{welling11}, and approximate message passing \cite{donoho09}—are formulated based on principles from statistical physics. These physics-inspired algorithms have not only enriched optimization theory and algorithms but also profoundly influenced computational physics, machine learning, and material and chemical design.

Gradient descent, one of the simplest first-order methods, has drawn much attention with the rise of machine learning, especially deep learning \cite{sutskever13}. Its low computational cost makes it suitable for large-scale and distributed optimization. Momentum based methods improve gradient descent by introducing a linear damping term \cite{POLYAK64}. Its accelerated variants, including Nesterov’s accelerated gradient descent (AGD) \cite{nesterov83} and adaptive methods such as Adam \cite{kingma2014adam}, further extend the momentum idea to improve convergence. From a physics perspective, a natural question arises: can nonlinear damping mechanisms achieve comparable or superior acceleration? Recent studies have incorporated dry friction \cite{adly22}, mixed viscous-Coulomb damping \cite{adly23}, and Hessian-dependent mechanisms, demonstrating promising performance in nonsmooth optimization.

However, to the best of our knowledge, despite the significant contributions of optimization theorists, including Attouch et al. \cite{attouch2022fast, adly22, adly23}, there are still many gaps in the explicit discussion of nonlinear damping effects. Specifically, the impact of a nonlinear kinetic energy term on the performance of related gradient descent algorithms has not been explored. This paper primarily addresses these gaps by providing a Hamiltonian framework that incorporates these effects. We propose a gradient-based optimization method with a nonlinear momentum mechanism, inspired by non-Newtonian mechanics, which achieves faster convergence than classical momentum methods while maintaining stability.

Inverse design problems in photonics \cite{Molesky18,Biehs21, ruan11, qiu12} often rely on adjoint-based gradient computations to efficiently calculate gradients for optimization. The proposed nonlinear momentum optimization method can be integrated with these adjoint-based approaches to enhance convergence speed and stability. Specifically, the proposed method utilizes gradient information to adjust the optimization trajectory with the flexibility of nonlinear momentum, thereby accelerating convergence and improving stability. Rather than directly computing the gradients, the method uses them to update design parameters effectively, making it particularly suitable for inverse photonic design tasks. This combination not only improves the efficiency of the optimization process but also provides a novel mechanism for tackling high-dimensional, non-convex photonic design problems. Results from numerical experiments validate the efficiency of the proposed method in photonics, suggesting its broader applicability to other fields, including machine learning.

The remainder of this paper is organized as follows. Section 2 introduces the dynamical modeling of momentum methods and presents the nonlinear damping construction and discretization. Section 3 provides a theoretical analysis of local convergence and computational aspects. Section 4 presents numerical experiments on representative nonconvex objectives and near-field thermal radiation design to validate the proposed method. Finally, Section 5 concludes the paper and outlines future work.

\section{Algorithm}
\subsection{Momentum Methods and Optimization Dynamics}
Momentum methods play a central role in both optimization theory and practice. The HB method \cite{kovachki21} treats the gradient descent update as the motion of a particle in a potential field, accelerating convergence by introducing a linear damping term. Within the framework of classical mechanics, momentum methods can be connected to the Euler–Lagrange equation :
\begin{eqnarray}
\frac{d}{dt} \Big( \frac{\partial L}{\partial \dot{x}_j}\Big) -  \frac{\partial L}{\partial x_j} = D_j \label{eq:1}
\end{eqnarray}
here $L$ is the Lagrangian, consisting of kinetic energy $K$ and potential energy $V$:
\begin{eqnarray}
L(x, \dot{x}, t) &=& K(\dot{x}) - V(x, t)
\end{eqnarray}
$D_j$ denotes generalized forces not derived from a potential, such as friction, which commonly appears in damped mechanical systems. In the case of linear damping, the frictional force is
proportional to the particle’s velocity, so the generalized force component in the x direction is given by:
\begin{eqnarray}
D_j = -\frac{\partial \mathcal{\phi}}{\partial \dot{x}_j} = -\gamma  \dot{x}(t) \label{eq:3}
\end{eqnarray}
where $\gamma$  is the damping coefficient, $\phi (\dot{x})$ is the dissipative function. Substituting  \eqref{eq:3} into \eqref{eq:1} yields the Lagrangian with a dissipative term:
\begin{eqnarray}
\frac{d}{dt} \Big( \frac{\partial L}{\partial \dot{x}_j}\Big) -  \frac{\partial L}{\partial x_j}  +\frac{\partial \mathcal{\phi}}{\partial \dot{x}_j} = 0,
\end{eqnarray}
The classical Heavy Ball method corresponds to linear damping, represented by the second-order ODE  \cite{kovachki21}:
\begin{eqnarray}
\ddot{x}(t) + \gamma \dot{x}(t) +\nabla V(x(t)) = 0.
\end{eqnarray}

\subsection{Anharmonic Kinetic Energy and Nonlinear Damping}
The aforementioned methods are fundamentally based on linear damping and Newtonian mechanics. However, in the context of nonconvex optimization, these traditional frameworks often suffer from persistent oscillations and reduced stability, leading to significant overshooting near extrema. In many complex physical systems \cite{bachtold22}—such as non-Newtonian fluids, viscoelastic materials, or oscillators with nonlinear damping—the inertial and damping behaviors are not well described by linear assumptions. Recently, several studies \cite{adly22, adly23} have modeled optimization dynamics as nonlinear dynamical systems to tackle more challenging objectives. Drawing inspiration from recent advances \cite{kardar2007, adly22, adly23}, we propose that incorporating generalized inertial and non-quadratic damping structures into gradient decent method can improve performance, particularly for nonconvex problems.

From the perspective of non-Newtonian mechanics, the Heavy Ball method can naturally be extended to a nonlinear momentum-based gradient descent method, described by the following nonlinear dynamical system:
\begin{eqnarray}
\ddot{x}(t) +\nabla \phi (\dot{x}(t))  + \nabla V(x(t)) = 0
\end{eqnarray}
This equation generalizes classical momentum method by extending the damping term to a velocity-dependent nonlinear dissipation function $\phi = \frac{\gamma}{\eta} ||\dot{x}||^{\eta}_{\eta} $. The parameter $\eta > 0$ controls the degree of nonlinearity: when $\eta = 2$ , the system reduces to linear damping (i.e., classical momentum); when $\eta > 0$, damping increases with velocity, consistent with the behavior of non-Newtonian media.

The nonlinear momentum optimization method proposed in this work is inspired by non-Newtonian mechanics, incorporating nonlinear damping terms and non-quadratic kinetic energy terms into the dynamical equations, thereby forming a class of nonlinear dissipative optimization dynamical systems. In the following, we focus primarily on the system’s physical dynamical behavior, energy dissipation structure, and algorithmic stability, providing a physical perspective for interpreting and qualitatively analyzing its optimization mechanisms. Rigorous and complete mathematical proofs of convergence are left to the field of optimization theory.

Consider a finite-dimensional mechanical system with generalized coordinate  $x\in \mathbb{R}^n$ and velocity $v=\dot{x}$. Let$K(v)$ be the kinetic energy function. Assume $K(v): \mathbb{R}^n\to \mathbb{R}$ is $\mathcal{C}^2$and strictly convex in $v$. And $V(x)$ be a $\mathcal{C}^1$ potential energy function; the dissipation (generalized damping) is given by a force $D(v)$ uch that $v\cdot D(v)\ge 0$for all $v$.

The equation of motion with dissipation is
\begin{eqnarray}
\frac{d}{dt} \Big( \nabla_v K(v) \Big) + \nabla_x V(x)  + D(v) = 0,
\end{eqnarray}

The dynamics (1) can be written in Hamiltonian form with dissipation, where the Hamiltonian is
\begin{eqnarray}
H(x, p):= K^*(p) + V(x) 
\end{eqnarray}
which equals the energy, $E(t) = H(x(t), p(t))$. The corresponding continuous-time optimization dynamics are then described by Hamilton’s equations:
\begin{eqnarray}
\dot{x}&=&\frac{\partial H}{\partial p} = v(p)   \\
\dot{p}&=& -\frac{\partial H}{\partial x} - D(v(p))= -\nabla_x V(x) - D(v)
\end{eqnarray}
This is equivalent to (1) under the Legendre duality.

Differentiate $H(x(t),p(t))$ with respect to $t$:
\begin{eqnarray}
\frac{d}{dt} H(x(t), p(t))  = \frac{\partial H}{\partial x}\cdot \dot{x} +\frac{\partial H}{\partial p}\cdot \dot{p}
\end{eqnarray}
Insert $\dot{x} = \partial_p H$ and $\dot{p} = - \partial_x H- D(v)$:
\begin{eqnarray}
 \frac{dH}{dt}  = \partial_x H \cdot \partial_p H  + \partial_p H \cdot (-\partial_x H - D(v))=-v \cdot D(v).
\end{eqnarray}
But $\partial_p H = \dot{x} = v$, Hence
\begin{eqnarray}
\frac{d E}{dt} = \frac{dH}{dt}  = -v \cdot D(v).
\end{eqnarray}

By the assumed dissipation criterion $v\cdot D(v)>0 $  , we obtain
\begin{eqnarray}
\frac{d}{dt}E \le 0.
\end{eqnarray}
If $v\cdot D(v) > 0$ for every nonzero $v$, then $E$ is strictly decreasing along any trajectory with $v\ne 0$, if $v\cdot D(v) =0 $ only at $v = 0$, then E is nonincreasing and can decrease until the system settles with $v=0$. That completes the proof under the stated assumptions. Let the dissipation force be

\begin{eqnarray}
D(v) = \gamma |v|^{\eta -1}  \odot \text{sgn}(v)
\end{eqnarray}
Where $|v|^{\eta-1}:= (|v_1|^{\eta-1}, |v_2|^{\eta-1}, ..., |v_n|^{\eta-1} )'$, and $\odot$ denotes elementwise product. Then
\begin{eqnarray}
v\cdot D(v) = \gamma \sum_i |v_i|^\eta = \gamma ||v||^{\eta}_{\eta} \ge 0.
\end{eqnarray}
so $dE/dt \le 0$.

Here, we consider non-quadratic kinetic energy function,
\begin{eqnarray}
K(v) = \frac{m^{s-1}}{s} \sum^{n}_{i=1} |v_i|^s = \frac{m^{s-1}}{s} ||v||^s_s, \qquad s > 1, \qquad m > 0
\end{eqnarray}
Compute the generalized momentum
\begin{eqnarray}
p_i = \frac{\partial K}{\partial v_i} =\frac{m^{s-1}}{s } s |v_i|^{s-1} \text{sgn}(v_i) = m^{s-1} |v_i|^{s-1}\text{sgn}(v_i)
\end{eqnarray}
Hence
\begin{eqnarray}
|p_i| = m^{s-1} |v_i|^{s-1} \Longrightarrow|v_i|^{s-1} = m^{1-s}|p_i| \Longrightarrow |v_i| = \frac{1}{m} |p_i|^{\frac{1}{s-1}}.
\end{eqnarray}
That is the mapping $v=\nabla_p K^*(p)$. Then,
\begin{eqnarray}
p_i v_i  =   \frac{1}{m} |p_i|^{\frac{s}{s-1}}
\end{eqnarray}
Also
\begin{eqnarray}
|v_i|^s  = \frac{1}{m^s}  |p_i|^{\frac{s}{s-1}}
\end{eqnarray}
So
\begin{eqnarray}
K(v) = \frac{m^{s-1}}{s} \sum^{n}_{i=1} |v_i|^s = \frac{m^{s-1}}{s} \sum_i \frac{1}{m^s} |p_i|^{\frac{s}{s-1}} =\frac{1}{s\cdot m} \sum_i |p_i|^{\frac{s}{s-1}}
\end{eqnarray}
Therefore,
\begin{eqnarray}
K^*(p) = \sum_i p_i v_i - K(v) = \frac{1}{m}(1-\frac{1}{s}) \sum^n_{i=1} |p_i|^{\frac{s}{s-1}}
\end{eqnarray}
Introduce the conjugate exponent  $r=\frac{s}{s-1}$ (so $1/s+1/r=1$). In vector norm form:
\begin{eqnarray}
K^*(p) = \frac{(s-1)}{s\cdot m} ||p||^r_r.
\end{eqnarray}

Differentiate $K^*$ with respect to $p_i$: 
\begin{eqnarray}
\frac{\partial K^*(p)}{\partial p_i} = \frac{(s-1)}{s\cdot m} r |p_i|^{r-1} \odot  \text{sgn}(p_i) = \frac{1}{m} |p_i|^{r-1} \odot  \text{sgn}(p_i).
\end{eqnarray}
Hence
\begin{eqnarray}
\partial_{p_i} K^*(p) = \frac{1}{m} \text{ sgn} (p_i) |p_i|^{\frac{1}{s-1}} = v_i,
\end{eqnarray}
which recovers the inversion $v = \nabla_p K^*(p)$ derived above.
The Hamiltonian is 
\begin{eqnarray}
H(x, p) = K^*(p) + V(x) = \frac{(s-1)}{s\cdot m} ||p||^r_r + V(x).
\end{eqnarray}

and we previously computed
\begin{eqnarray}
\frac{dH}{dt}  = -v \cdot D(v).
\end{eqnarray}
And the monotone decay of the Hamiltonian follows since $v \cdot D(v) \ge 0$.

It should be noted that in conventional gradient based optimization dynamics, the kinetic energy term is quadratic, corresponding to linear velocity dependence in a Newtonian system. In this work, the restriction can be lifted by generalizing the kinetic energy to:
\begin{eqnarray}
K(\dot{x}) &=& \frac{m^{s-1}}{s} ||\dot{x}||^{s}_s 
\end{eqnarray}
where $m$ is the particle mass. While alternative kinetic energy formulations are possible, we adopt the power-law form in this study.

\subsection{Discretization and Update Scheme}
To apply the continuous-time dynamics to iterative optimization, we discretize the Hamilton equations. Let $x_k\approx x(t_k)$, $p_k\approx p(t_k)$ and set $m = 1$. The Heavy Ball method naturally arises as:
\begin{eqnarray}
p_k&=& - h \nabla V(x_k ) +(1- \gamma h) p_{k-1}\\
x_{k+1}&=&x_k + h p_k
\end{eqnarray}
Let $\alpha = h^2$, $\beta = 1-\gamma h$, and $hp_k \to p_k$, we recover the standard Heavy Ball update. Replacing $\nabla V(x_k) \to \nabla V(x_k + \beta p_{k-1})$, we get Nesterov's accelerated gradient method.
\begin{eqnarray}
x_{k+1}&=&x_k -\alpha \nabla V(x_k + \beta (x_k - x_{k-1})) +  \beta (x_k - x_{k-1})
\end{eqnarray}
For the proposed algorithm, the update rules are:
\begin{eqnarray}
p_k&=& p_{k-1} - h \nabla V(x_k ) - h \nabla \phi(p_{k-1})\\
x_{k+1}&=&x_k +  h \nabla K^* (p_{k})   
\end{eqnarray}
here $\phi(p_k)$ controls the damping, and $K^*(p_k)$ determines the anharmonic kinetic energy. In the update scheme above, $\gamma>0 $ is the nonlinear damping coefficient,  $\eta$ controls the nonlinearity of the damping, and $h>0$ is the step size (time interval or learning rate). The basic nonlinear momentum algorithm for minimizing $V(x)$ is given in Algorithm \ref{alg:alg1}.
\begin{algorithm}
\caption{}\label{alg:alg1}
\begin{algorithmic}[1]
\State \textbf{input:}  initial $x_0$,  parameters $h, \gamma, \eta, r, n_{max}$
\State \textbf{initialize:} $p_{-1} = 0$ 
\State \textbf{for} $k$ \textbf{in} $0:n_{max}$
\State  \qquad $p_k \leftarrow  p_{k-1} - h \nabla V(x_k ) - h \nabla \phi(p_{k-1}),  \qquad \triangleleft  \quad \phi = \frac{\gamma}{\eta} ||p||^{\eta}_{\eta} $
\State  \qquad $x_{k+1} \leftarrow x_k +  h \nabla K^* (p_{k})   \qquad \triangleleft  \quad K^* = \frac{1}{r} ||p||^{r}_{r} $
\State \textbf{return} $x_{n_{max}}$
\end{algorithmic}
\end{algorithm}

The discrete updates emulate the dynamics of a velocity-dependent nonlinear dissipative system, with the following properties: when $\eta = 2$, the algorithm reduces to the classical Heavy Ball method; when $\eta = 1$, damping scales with the velocity magnitude, effectively suppressing high-speed oscillations; when $1< \eta < 2$, momentum is retained in low-speed regions, helping the system escape saddle points or flat plateaus. Furthermore, Nesterov's AGD with nonlinear momentum for minimizing $V(x)$ is given in Algorithm \ref{alg:alg2}.
\begin{algorithm}
\caption{}\label{alg:alg2}
\begin{algorithmic}[1]
\State \textbf{input:}  initial $x_0$,  parameters $h, \gamma, \eta, r, n_{max}$
\State \textbf{initialize:} $p_{-1} = 0$ 
\State \textbf{for} $k$ \textbf{in} $0:n_{max}$
\State  \qquad $y_{k} \leftarrow x_k +   p_{k-1} - \gamma h \phi(p_{k-1})  $
\State  \qquad $p_k \leftarrow  p_{k-1} - h \nabla V(y_k ) - h \nabla \phi(p_{k-1}) $
\State  \qquad $x_{k+1} \leftarrow x_k +  h \nabla K^* (p_{k})  $
\State \textbf{return} $x_{n_{max}}$
\end{algorithmic}
\end{algorithm}

It should be noted that both the proposed nonlinear momentum method and mirror descent \cite{nemirovskiyudin1983} reinterpret gradient-based optimization from a generalized dynamical or geometric perspective, they differ in their underlying principles and objectives. Mirror descent reformulates the optimization process in a non-Euclidean geometry defined by a mirror map or a Bregman divergence. It effectively performs gradient descent in the dual space, adapting the update direction and step size according to the local curvature or information geometry of the problem. The goal is to ensure better alignment between the optimization geometry and the structure of the objective function.

In contrast, the nonlinear momentum method extends the temporal dynamics rather than the spatial geometry of optimization. It originates from the analogy between gradient methods and physical systems governed by kinetic energy and damping. By introducing an anharmonic (non-quadratic) kinetic energy and nonlinear damping, the method generalizes classical momentum schemes to capture richer inertial behaviors and adaptive momentum attenuation. Thus, while mirror descent adapts the geometry of the descent trajectory, the nonlinear momentum method adapts its dynamical law.

Despite this distinction, the two frameworks are conceptually connected. Both can be viewed as geometric or dynamical regularizations of gradient descent: mirror descent modifies the metric tensor of the optimization manifold, whereas the nonlinear momentum method modifies the kinetic energy metric of the underlying dynamical system. In principle, a unified interpretation can be achieved by formulating the nonlinear momentum dynamics in a Riemannian space defined by a mirror map, suggesting a possible bridge between the two approaches.

Since mirror descent can naturally handle constraints and possesses better convergence and stability properties, we propose to combine the nonlinear momentum mechanism with the mirror descent framework. This yields a generalized nonlinear momentum algorithm, which integrates the advantages of adaptive momentum dynamics and geometry-aware updates.
\begin{eqnarray}
p_k&=& p_{k-1} - h \nabla V(x_k ) - h \nabla \phi(p_{k-1})\\
z_{k+1}&=&\nabla \psi(x_k) +  h \nabla K^* (p_{k})   \\
x_{k+1}&=&\nabla \psi^*(z_{k+1})
\end{eqnarray}
where $\psi(\cdot)$ is the mirror potential function and $\psi^*(\cdot)$ its conjugate. For example, in the $l_p$ geometry, $\psi(z) = \frac{1}{p}\sum |z_i|^p$ and  $\psi^*(x) = \frac{1}{q}\sum |x_i|^q$ ( $\frac{1}{p} + \frac{1}{q} = 1$).  Mirror descent with nonlinear momentum for minimizing $V(x)$ is given in Algorithm \ref{alg:alg3}.

\begin{algorithm}
\caption{}\label{alg:alg3}
\begin{algorithmic}[1]
\State \textbf{input:}  initial $x_0$,  parameters $h, \gamma, \eta, r,  q, n_{max}$
\State \textbf{initialize:} $p_{-1} = 0$ 
\State \textbf{for} $k$ \textbf{in} $0:n_{max}$
\State  \qquad $p_k \leftarrow  p_{k-1} - h \nabla V(y_k ) - h \nabla \phi(p_{k-1}) $
\State  \qquad $z_{k+1} \leftarrow \nabla \psi(x_k) +  h \nabla K^* (p_{k})  $
\State  \qquad $x_{k+1} \leftarrow \nabla \psi^*(z_{k+1})  $
\State \textbf{return} $x_{n_{max}}$
\end{algorithmic}
\end{algorithm}

\section{Convergence analysis}
In this section we perform a convergence analysis for the algorithm with quadratic kinetic energy with nonlinear damping. We assume objective function $V(x)$ to be $m$-strongly convex and $L$-smooth. $L$-smoothness, means that its gradient is Lipschitz continuous:
\begin{eqnarray}
||\nabla V(x) -\nabla V(y) || \le L || x- y||
\end{eqnarray}
where L > 0 is the Lipschitz constant.This further leads to the following descent lemma, which is fundamental in optimization:
\begin{eqnarray}
V(y) \le V(x) + \langle \nabla V(x), y - x \rangle+ \frac{L}{2} ||y - x||^2
\end{eqnarray}
We define the energy function,
\begin{eqnarray}
 E_k = \frac{m}{2}||v_k||^2 + V(x_k) 
\end{eqnarray}
Then, the successive energy difference can then be written as
\begin{eqnarray}
E_{k+1} - E_k = V(x_{k+1}) - V(x_k) + \frac{m}{2} (||v_{k+1}||^2 - ||v_k||^2)
\end{eqnarray}
By the descent lemma,
\begin{eqnarray}
V(x_{k+1}) \le V(x_k) + \langle \nabla V(x_k), x_{k+1} - x_k\rangle + \frac{L}{2} ||x_{k+1} - x_k||^2
\end{eqnarray}
Substituting $v_{k+1} = x_{k+1}-x_k = v_k - h \nabla V(x_k) - \gamma h |v_k|^{\eta-1} \odot \text{sgn} (v_k)$. Let $v_k^{\eta-1} := |v_k|^{\eta-1} \odot \text{sgn} (v_k)$, we obtain
\begin{eqnarray}
V(x_{k+1}) \le V(x_k) + \langle \nabla V(x_k), v_k \rangle -  h ||\nabla V(x_k)||^2 -  \gamma h \langle \nabla V(x_k), v_k^{\eta-1} \rangle + \frac{L}{2} ||v_{k+1}||^2
\end{eqnarray}
With the identity,
\begin{eqnarray}
\langle \nabla V(x_k), v_k -  \gamma h v_k^{\eta-1}  \rangle =\frac{1}{2h} || v_k -  \gamma h v_k^{\eta-1} ||^2 + \frac{h}{2} || \nabla V(x_k)||^2 -  \frac{1}{2h} ||v_{k+1}||^2 
\end{eqnarray}
We obtain,
\begin{eqnarray}
\Delta E \le  - \frac{h}{2} ||\nabla V(x_k)||^2 + \Big(\frac{L}{2}  -\frac{1}{2h} + \frac{m}{2}\Big)  ||v_{k+1}||^2 +  \frac{1}{2h} || v_k -  \gamma h v_k^{\eta-1} ||^2 -  \frac{m}{2}||v_{k}||^2
\end{eqnarray}
When $|v_{k}| < 1$,  $1< \eta \le 2$, we have
\begin{eqnarray}
(1-\gamma h)^2 || v_k^{\eta-1} ||^2   \le || v_k -  \gamma h v_k^{\eta-1} ||^2 \le   (1-\gamma h)^2 || v_k ||^2
\end{eqnarray}
\begin{eqnarray}
\Delta E \le -\frac{h}{2}  ||\nabla V(x_k)||^2 + \Big(\frac{L}{2}  -\frac{1}{2h} + \frac{m}{2}\Big)  ||v_{k+1}||^2 + \Big( \frac{(1-\gamma h)^2}{2h}  -  \frac{m}{2} \Big)||v_{k}||^2
\end{eqnarray}
If the parameters,
\begin{eqnarray}
(L  + m) h &\le& 1\\
(1-\gamma h)^2  -  m h &\le& 0
\end{eqnarray}
We get
\begin{eqnarray}
\Delta E \le 0
\end{eqnarray}

Based on asymptotic convergence analysis, \cite{attouch2022fast} proves that the case $\eta = 2$ separates weak damping ($\eta > 2$) from strong damping ($\eta < 2$). For a more mathematically rigorous treatment, readers may refer to this paper.

\section{Numerical experiments}

To validate the effectiveness and stability of the proposed nonlinear momentum optimization method, this section presents several numerical experiments comparing the convergence performance of different momentum mechanisms on representative optimization tasks, with a particular focus on the adaptability of nonlinear damping and non-quadratic kinetic energy structures in complex optimization problems.

\subsection{Rosenbrock Function}
As the first example, we consider the Rosenbrock function, a widely used benchmark nonconvex test function in optimization:

\begin{eqnarray}
V(x) = \sum^{d-1}_{i=1}[100(x_{i+1} - x^2_i)^2 - (x_i - 1)^2]
\end{eqnarray}

\begin{figure}[h!]
\centering
\begin{subfigure}
   \centering
    \includegraphics[width=.5\textwidth]{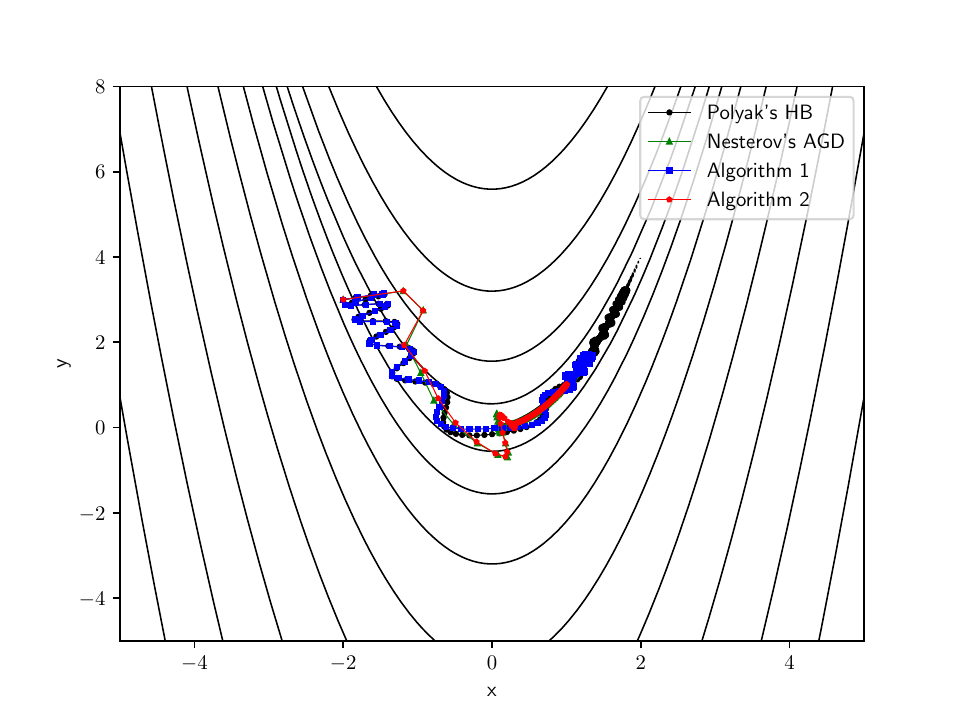}
\end{subfigure}
\hfill
\begin{subfigure}
   \centering
    \includegraphics[width=0.5\textwidth]{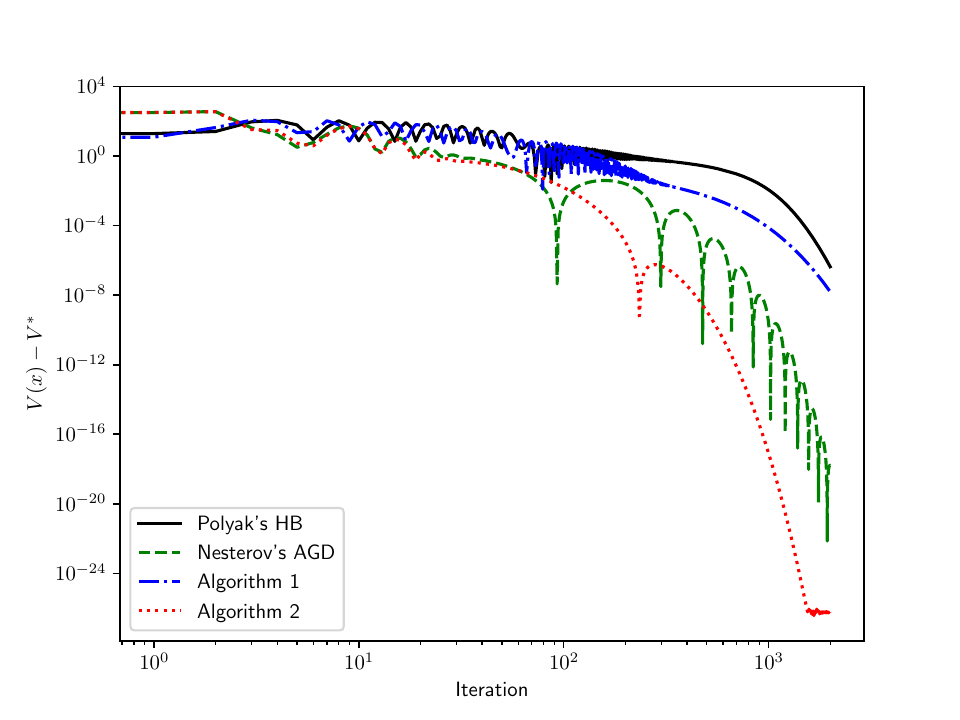}
\end{subfigure}
\caption{Comparisons between HB, AGD, and Algorithm 1 and Algorithm
2 for the Rosenbrock objective function. The learning rate $h$ for HB and Algorithm 1 (with $\eta = s = 1.9$) is 0.0002, while for AGD and Algorithm 2 (with $\eta = s = 1.98$), the learning rate $h$ is 0.001. And $\gamma h= 0.02$, $x_0 = (-2.0, 3.0)'$ for all these algorithms.}
\label{fig:fig1}
\end{figure}

The Rosenbrock function features a highly curved valley, which can be interpreted as a special potential energy landscape, making it suitable for evaluating algorithmic convergence under nonlinear coupling. In this task, we compare the following optimization algorithms: the classical Heavy Ball (HB) method, representing a momentum mechanism with linear damping and quadratic kinetic energy; the proposed nonlinear momentum method; and Nesterov’s Accelerated Gradient Decent method. s Figure \ref{fig:fig1} shows for the  Rosenbrock objective function, Algorithm 1 and Algorithm 2 demonstrate competitive performance against HB and AGD, outperforming them in this case.

\begin{figure}
\centering
\begin{subfigure}
   \centering
    \includegraphics[width=.5\textwidth]{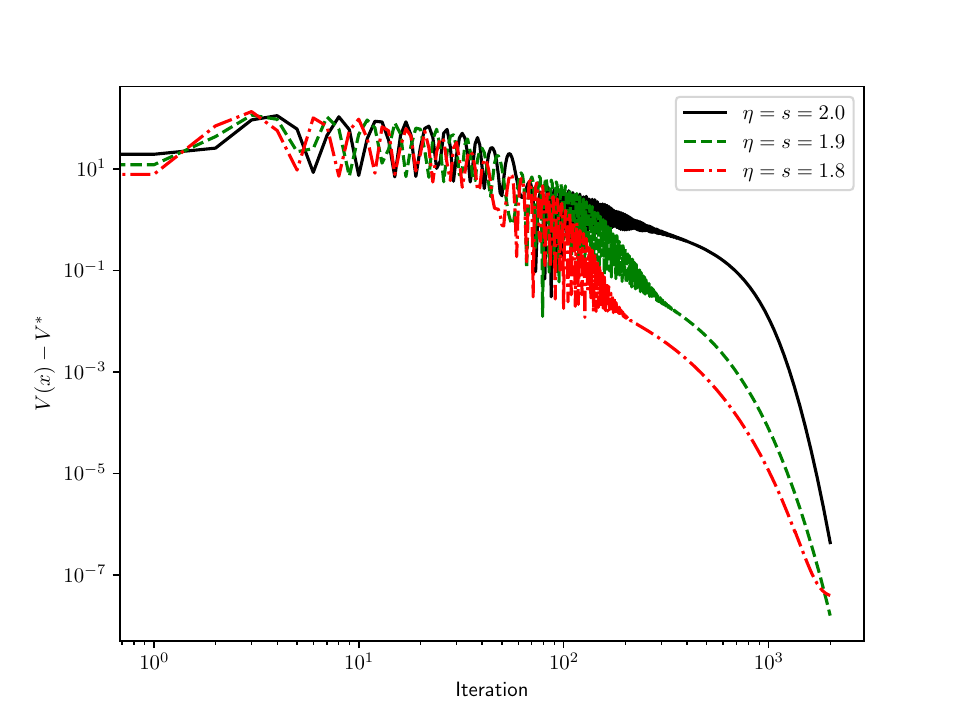}
\end{subfigure}
\hfill
\begin{subfigure}
   \centering
    \includegraphics[width=0.5\textwidth]{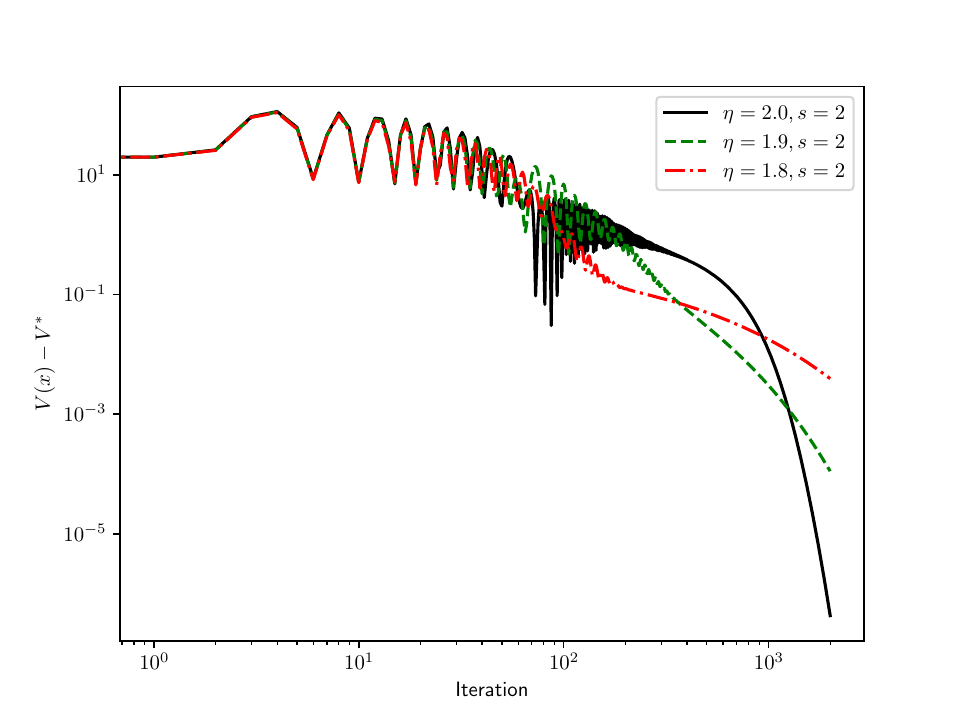}
\end{subfigure}
\hfill
\begin{subfigure}
   \centering
    \includegraphics[width=0.5\textwidth]{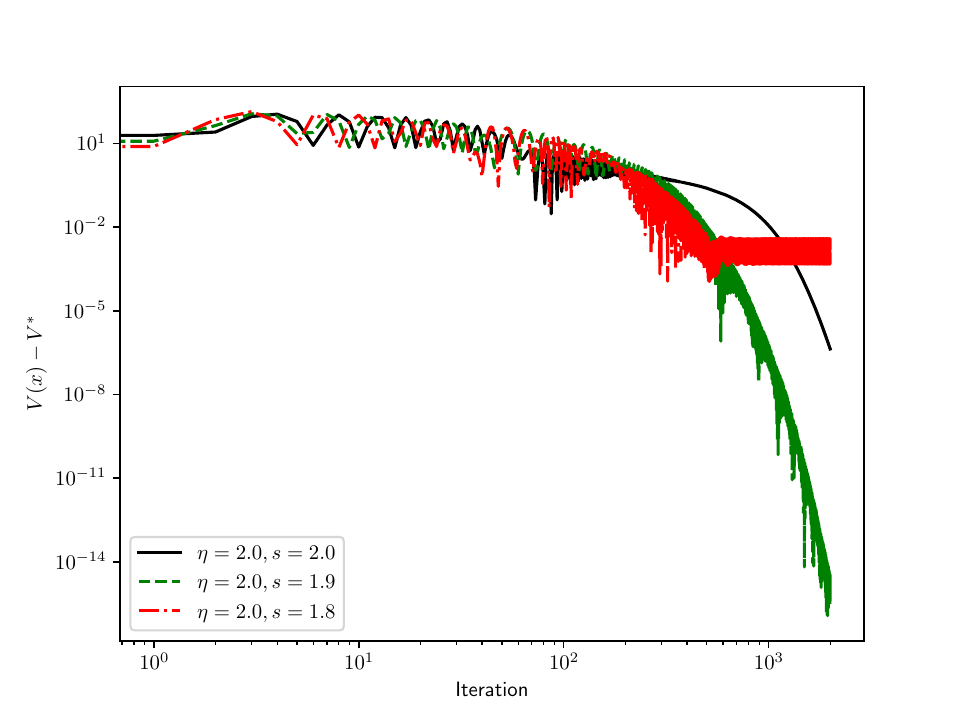}
\end{subfigure}
\caption{Comparisons between different nonlinear orders for kinetic energy and dissipative potential used in Algorithm 1. The learning $h = 0.0002$, damping coefficient $\gamma h= 0.02$ for all these simulations.  }
\label{fig:fig2}
\end{figure}

\subsection{Inverse Design of Multilayer Nanospheres}
In the 2nd example, we consider an inverse design problem for a multilayer nanosphere composed of alternating silver ($Ag$) and silica ($SiO_2$) shells. The objective is to optimize the layer thicknesses so that the resulting structure exhibits high average emissivity within the visible range (400-800 nm). The design variable is the set of shell thicknesses
\begin{eqnarray}
x = (d_1, d_2,..., d_N)',
\end{eqnarray}
where N is the number of layers. Physical constraints include manufacturability (minimum layer thickness >5nm) and an upper bound on the overall radius 
$R_{\max} = 300nm$

\begin{figure}
\centering
\includegraphics[width=0.7\linewidth]{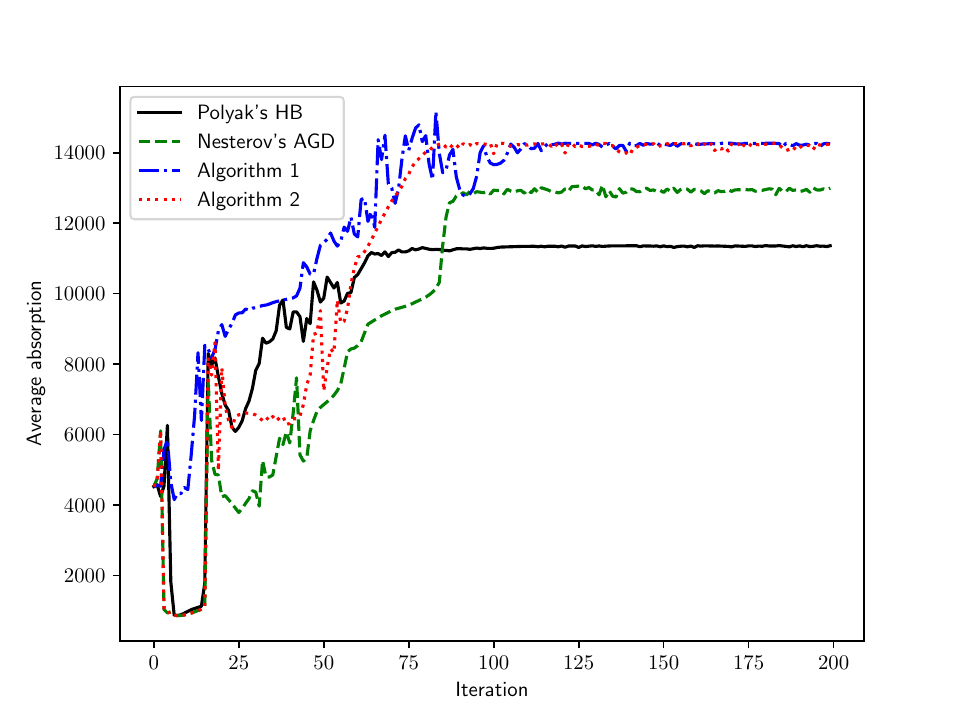}
\caption{HB, AGD, and Algorithm 1 and Algorithm 2 for the numerical example 2. The learning rate $h$ 0.001, $\gamma h= 0.1$ for all these algorithms. And $\eta = s = 1.95$ for Algorithm 1 and Algorithm 2}
\label{fig:fig3}
\end{figure}

\begin{figure}
\centering
\includegraphics[width=0.7\linewidth]{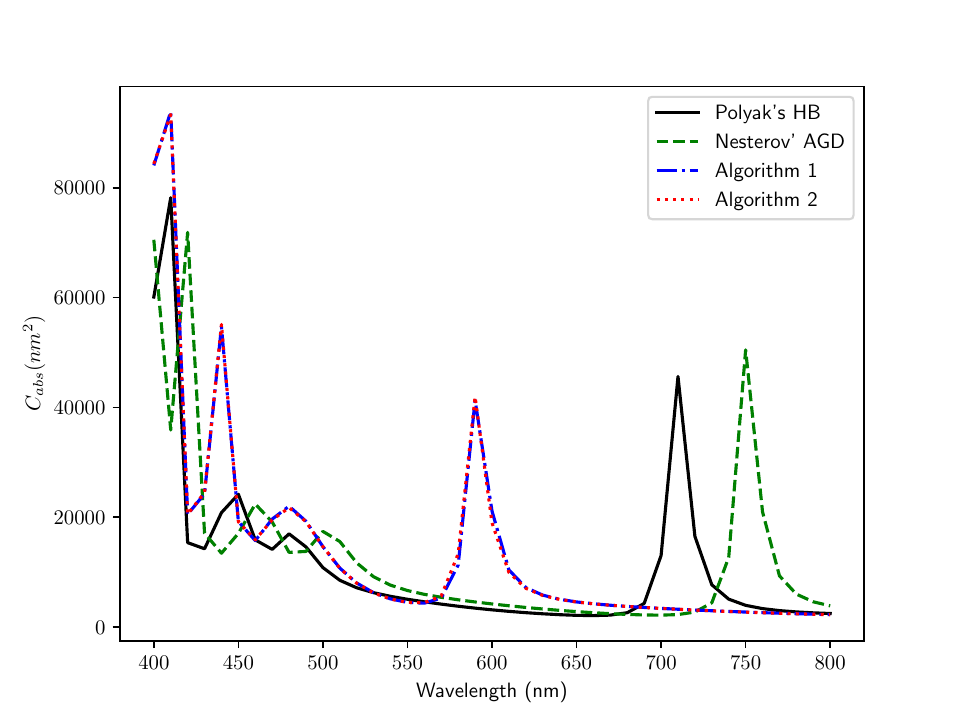}
\caption{Average absorption cross-sections over visible wavelength range. The method for computing the absorption cross section is mRTMA \cite{zhang2025}.}
\label{fig:fig4}
\end{figure}

The spectral absorption efficiency $Q_{abs}(\lambda; x)$ of the multilayer sphere is computed using the modified recursive transfer matrix algorithm (mRTMA), a method previously developed by \cite{zhang2025} for concentric layered spheres. The broadband objective is defined as the spectral average:
\begin{eqnarray}
J(x) = \frac{1}{\lambda_2 - \lambda_1} \int^{\lambda_2}_{\lambda_1} \sigma_{abs}(\lambda; x) d\lambda, 
\end{eqnarray}
where $\lambda_1 = 400nm, \lambda_2 = 800nm$. And we are concerned with the weighted average absorptivity, which is defined as the arithmetic mean. The design is subject to the following layer thickness bounds:
\begin{eqnarray}
d_{min}\le d\le d_{max}
\end{eqnarray}
where the lower bound ensures the validity of the effective continuous dielectric description and the upper bound reflects fabrication limits. The overall particle radius must not exceed a prescribed maximum value. For silver layers, the minimum thickness must be sufficiently large to form a continuous film. For $N=3$ layers,  the optimal design exhibits a balance between plasmonic resonance enhancement from $Ag$ shells and dielectric confinement from $SiO_2$ layers. Figure \ref{fig:fig3} and Figure \ref{fig:fig4} present the results. We can see that the Algorithm 1 and Algorithm 2 converge faster than HB, and AGD. Figure \ref{fig:fig4} also indicates that there are many local optima for the average absorption cross-section objective, which are related to different plasmonic resonance absorption spectrum structures for the multilayer nanosphere.

It should be noted that the results obtained in this numerical example correspond to a local optimum only. In practical inverse photonic design problems, identifying the global optimum of design parameters typically requires the use of global optimization strategies, such as the Multi-Level Single-Linkage (MLSL) algorithm, which selects the best solution among multiple local optima.

\section{Conclusion}

Based on non-Newtonian dynamics, this paper proposes a novel nonlinear momentum optimization method. The method significantly improves the convergence speed of momentum-based gradient optimization algorithms by generalizing the kinetic energy formulation and introducing a nonlinear damping mechanism. Specifically, an anharmonic kinetic energy function is used to capture the inertial effects of accumulated gradient information during iterations, while the nonlinear damping mechanism adaptively regulates the influence of the momentum term on the convergence trajectory. Numerical experiments show that the proposed algorithm offers significant advantages in terms of parameter flexibility, stability, and convergence speed compared to classical momentum-based methods. In particular, the method outperforms traditional gradient methods in inverse design tasks. Additionally, this method provides a new physics-based perspective on numerical optimization analysis, further expanding the application of physics-inspired optimization algorithms.

This study not only demonstrates significant performance improvements in inverse design tasks but also opens new paths for the development of physics-inspired optimization algorithms. Future research directions include: employing higher-order numerical integration schemes, exploring more general forms of kinetic energy, and combining advanced sampling and optimization strategies within the Hamiltonian Monte Carlo framework to further enhance efficiency. Additionally, systematic theoretical analysis is needed to deepen the understanding of the convergence and stability properties of nonlinear momentum mechanisms.

\section{Acknowledgements}

\bibliographystyle{unsrt}  
\bibliography{reference}  


\end{document}